\PassOptionsToPackage{table}{xcolor}

\documentclass[sigconf]{acmart}

\usepackage{soul}
\usepackage{algorithm}
\usepackage{algorithmic}
\usepackage{multirow}
\usepackage{makecell}
\usepackage{colortbl}
\usepackage{pifont}          
\usepackage[most]{tcolorbox}
\usepackage{wrapfig}
\usepackage{subcaption}
\usepackage{placeins}        

\definecolor{lightblue}{RGB}{235, 245, 255}

\DeclareMathOperator*{\argmax}{argmax}

\setcopyright{acmlicensed}
\copyrightyear{2026}
\acmYear{2026}
\acmDOI{XXXXXXX.XXXXXXX}
\acmConference[CIKM '26]{The 35th ACM International Conference on
  Information and Knowledge Management}{November 7-11, 2026}{Rome, Italy}
\acmISBN{978-1-4503-XXXX-X/26/11}

\title[Tutoring Effectiveness Index]{The Tutoring Effectiveness Index:
  Predicting LLM Math Tutor Quality from Four Conversation Signals}

\author{Shim Jaechang}
\email{wockd9799@gmail.com}
\affiliation{%
  \institution{Chosun University}
  \country{South Korea}
}

\author{Unggi Lee}
\authornote{Corresponding author.}
\email{codingchild@korea.ac.kr}
\affiliation{%
  \institution{Korea University Sejong Campus}
  \country{South Korea}
}

\renewcommand{\shortauthors}{Shim and Lee}

\begin{document}

\begin{abstract}

Aligning large language models (LLMs) as math tutors typically demands costly reinforcement-learning (RL) training and external LLM judges. We ask whether a frozen model's internal reasoning signals can replace both. We propose the Tutoring Effectiveness Index (TEI), a training-free, judge-free four-signal index that combines a Schoenfeld-Verify keyword ratio, a math-step density, an ends-question rate, and a deep-reasoning gate from the Deep-Thinking Ratio (DTR) probe. Selecting from $N$ candidates with TEI (the TEI@$N$ rule) raises the improvement rate on pre-incorrect scenarios from $59.0\%$ to $81.9\%$ at $N{=}8$ on a frozen DeepSeek-R1-8B base, with no training and no external judge. We also measure the alignment tax of pedagogical GRPO. Thinking length drops from $1{,}764$ to $119$ words per turn ($-93\%$), Content-Knowledge and Pedagogical-Knowledge accuracy fall by $-71\%$ and $-80\%$ relative, and the student's $\Delta$ Solve Rate crosses from $+0.180$ to $-0.012$. To anchor the behavioural reading, we reproduce an 82-code educational codebook on $119{,}009$ tutor sentences with a one-shot structural classifier. Together, these results offer a cost-effective recipe for building math-tutoring LLMs without RL training or external judges.

\end{abstract}



\maketitle

\section{Introduction}
\label{sec:intro}

A reasoning LLM that passes a math benchmark does not yet pass as a tutor. Pedagogical alignment, making the model \emph{teach} rather than merely solve, is currently approached by reinforcement learning (RL) over pedagogical rewards~\cite{dinucujianu2025pedagogicalrl, lee2026pedagogicalrl-thinking, learnlmteam2025learnlm}, which requires multi-GPU training and curated data, and evaluated by LLM-as-judge protocols~\cite{zheng2023llmjudge, tack2022aiteacher}, which carry same-model-bias risk. Both share an unstated assumption that improving tutoring requires \emph{changing} the model. This paper takes the opposite stance. We keep the model frozen, add no external judge at selection time, and ask whether the LLM's own internal signals, already available during decoding, are enough to pick a better tutor response from $N$ candidates.

A natural first attempt borrows from the reasoning literature. \citet{chen2026dtr} show that the Deep-Thinking Ratio (DTR), a token-level probe of cross-layer deliberation, robustly predicts mathematical accuracy ($r = 0.683$). We find that it fails on tutoring (all $|r|<0.09$, and a causal steering intervention that raises DTR by $17.3\%$ confirms DTR alone is uninformative about tutoring quality). After combining multiple conversation-log signals and testing them jointly against learning outcomes, we propose the Tutoring Effectiveness Index (\emph{TEI}). It combines four signals computed without any external model at selection time. These are a verification ratio extracted from the thinking trace via a Schoenfeld-Verify keyword regex, a math-step density and an ends-question rate from the visible output, and a deep-reasoning gate read off the DTR probe. The four signals use fixed weights, and the per-token DTR probe runs as a forward hook on the decoding pass. \emph{TEI@$n$} generates $N$ candidates and selects the highest-scoring one.

We evaluate on a multi-turn in-domain tutoring set built from BigMath~\cite{albalak2025bigmath} and on the out-of-distribution (OOD) OpenLearnLM suite~\cite{lee2026openlearnlm}. Three patterns emerge. First, Group Relative Policy Optimization (GRPO) pedagogical alignment of a reasoning base model lowers actual learning ($\Delta$ Solve Rate from $+.180$ to $-.012$), compresses the thinking trace by an order of magnitude, and damages OOD mathematical knowledge. Second, \emph{TEI@$n$} selection on the base model raises the improvement rate on pre-incorrect scenarios from $59.0\%$ to $81.9\%$ at $N=8$, at no training cost. Third, DTR-based selection improves over Greedy by only $7$ to $14$ percentage points across $N$, mirroring its zero correlation with quality. A behavioural-codebook analysis on $119{,}009$ tutor sentences (Section~\ref{sec:analysis}) traces the alignment shift to a collapse in Mathematical Knowledge for Teaching utterances and a surge in Socratic-style Exploratory questions, confirming that the pedagogical RL reward steers the policy toward a proxy decoupled from learning.

The behavioural analysis is itself part of the contribution. By inlining an 82-code educational codebook into a single GPT-4o-mini prompt at temperature $0$, we re-label every tutor sentence in our corpus and use those labels for the four-category and per-turn analyses in Section~\ref{sec:analysis}. The combined picture gives deployers a measurable warning against unconditional pedagogical RL and a lightweight, inspectable alternative they can apply to any frozen reasoning LLM without retraining or external judges.

\subsection{Contributions}
\label{sec:intro:contributions}

\begin{itemize}
\item \emph{C1.} \emph{TEI}, a training-free, judge-free four-signal index (Schoenfeld-Verify regex, math-step density, ends-question rate, DTR gate).
\item \emph{C2.} \emph{TEI@$n$}, an inference-time selector that beats Greedy by $+22.9$pp at $N=8$ on DeepSeek-R1-8B base, with no training.
\item \emph{C3.} A one-shot reproduction of an 82-code educational codebook on $119{,}009$ tutor sentences.
\item \emph{C4.} A multi-dimensional alignment tax in pedagogical RL ($\Delta$Sol $-0.192$, thinking length $-93\%$, OL CK/PK $-71\%$/$-80\%$ relative).
\end{itemize}

\section{Related Work}
\label{sec:related}

Pedagogically-aligned LLM tutors are developed primarily through RL over pedagogical rewards. PedagogicalRL~\cite{dinucujianu2025pedagogicalrl} and its thinking-augmented successor~\cite{lee2026pedagogicalrl-thinking} train tutoring policies with GRPO over combinations of solve rate, leak avoidance, helpfulness, and thinking-trace quality; LearnLM~\cite{learnlmteam2025learnlm} aligns Gemini through human pedagogical feedback. These methods produce strong tutors on average but require GPU clusters, curated corpora, and re-training whenever the reward composition changes. Evaluation benchmarks include MathDial~\cite{macina2023mathdial}, OpenLearnLM~\cite{lee2026openlearnlm}, and the earlier conversational study of~\citet{tack2022aiteacher}. We shift pedagogical alignment from training to inference, leaving the model frozen and exposing the alignment lever as an interpretable selection rule over candidate responses.

On the inference-time side, prior work scales frozen LLMs through chain-of-thought prompting~\cite{wei2022cot}, self-consistency~\cite{wang2023selfconsistency}, internal probes of deliberation depth such as DTR~\cite{chen2026dtr}, and activation steering~\cite{turner2023activationsteering}. DTR predicts mathematical correctness strongly in its original setting but does not transfer to the educational dialogue setting (Section~\ref{sec:result:rq1}). The proxy our index does use is instead grounded in a long tradition of behavioural coding of tutor-student interaction: Schoenfeld episodes~\cite{schoenfeld1985mps}, Polya phases~\cite{polya1945howto}, MKT~\cite{shulman1986knowledge, ball2008contentknowledge}, scaffolding~\cite{wood1976tutoring}, the Interactive-Constructive-Active-Passive (ICAP) framework~\cite{chi2014icap}, and Socratic dialogue~\cite{liu2024socraticlm}. We adopt the 82-code educational codebook of~\citet{lee2026pedagogicalrl-thinking} for our behavioural analysis. To our knowledge, ours is the first work to combine hidden-state reasoning probes with surface dialogue features into a unified judge-free selection criterion for tutor responses; same-model-bias concerns about LLM-as-judge~\cite{zheng2023llmjudge} motivate this judge-free design.

\section{Method}
\label{sec:method}

\subsection{Problem Formulation}
\label{sec:method:formulation}

Given a tutoring context $x$ and a frozen reasoning LLM $p_\theta$, our goal is to select a single response from $N$ stochastic candidates $\{y_1, \dots, y_N\} \sim p_\theta(\cdot \mid x)$ that improves a held-out learning outcome, without (i) additional training on $\theta$ and (ii) external reward or judge models at selection time. We formalise this as choosing a scoring function $s: \mathcal{Y} \to \mathbb{R}$ such that the predicted-best candidate $y^\star = \argmax_i s(y_i)$ improves expected $\Delta(y, x)$. The two constraints distinguish our setting from RL alignment (which retrains $\theta$) and from judge-based test-time selection (which queries a separate model per candidate).

\subsection{Tutoring Effectiveness Index (TEI)}
\label{sec:method:tei}

DTR captures \emph{how} a model reasons, but not \emph{what} kind of reasoning is being done. Effective tutoring requires not just deep computation but a particular interaction pattern. The tutor should verify student understanding, sustain problem-relevant guidance, and avoid answer disclosure. We borrow the verification idea from Schoenfeld's episode model of mathematical problem solving~\cite{schoenfeld1985mps}, where \emph{Verify} marks the moments a solver pauses to check intermediate work, and from the GRPO-tuned tutoring policies of~\citet{lee2026pedagogicalrl-thinking}, which already reward verification-rich thinking traces (we reuse their 82-code codebook for the behavioural analysis in Section~\ref{sec:analysis}).

We extract four signals from the conversation log (one from the model's internal thinking trace, two from the visible tutor output, and one from the layer-wise reasoning probe), each grounded in a distinct theoretical anchor (Table~\ref{tab:tei-signals}).

\begin{table}[h]
\small
\centering
\caption{\emph{TEI} signals: one thinking-trace marker, two surface markers, and one deep-reasoning gate. All four are computed without external models at selection time.}
\label{tab:tei-signals}
\begin{tabular}{@{}lll@{}}
\toprule
Signal & Source & Theoretical anchor \\
\midrule
$V$: verify ratio    & regex on thinking trace & Schoenfeld Verify episode \\
$\tilde M$: math steps & regex on visible output & direct mathematical guidance \\
$Q$: ends-question   & regex on visible output  & Socratic over-questioning \\
$D = \mathbf{1}[\mathrm{DTR}\!\geq\!0.4]$ & per-token JSD probe & deep-reasoning gate \\
\bottomrule
\end{tabular}
\end{table}

The Tutoring Effectiveness Index combines the four signals with fixed weights:
\begin{equation}
\mathrm{TEI}(y) \;=\; \alpha \, V(y) + \beta \, \tilde M(y) - \gamma \, Q(y) + \delta \, \mathbf{1}[\mathrm{DTR}(y) \geq 0.4],
\label{eq:tei}
\end{equation}
with $(\alpha, \beta, \gamma, \delta) = (1.0, 0.75, 1.0, 0.5)$ and $\tilde M$ min-max normalised within the candidate pool (with a denominator cap of $20$ math-step matches per turn). The DTR probe used by the gate $D$~\cite{chen2026dtr} is computed in log space via forward hooks on each decoder layer. It fires when at least $40\%$ of generated tokens remain layer-wise unsettled below a JSD threshold $g=0.5$ until depth $\rho \cdot L$ with $\rho = 0.85$. The signs of the four weights, not their magnitudes, carry the substantive claim. Verification and explicit math content are rewarded, closing tutor turns with questions is penalised, and the deep-reasoning gate adds a small bonus when the DTR probe indicates sustained layer-wise deliberation.

\paragraph{Why fixed weights.} Calibrating weights against a labelled outcome would require either external judges or learning-gain probes, undermining the training-free, judge-free property. Eq.~\eqref{eq:tei} therefore encodes three priors only: (i) verification is rewarded; (ii) explicit math guidance contributes positively; (iii) ending tutor turns with questions, a documented Socratic over-pattern in math tutoring studies, is penalised. A constrained grid sweep around these weights (Section~\ref{sec:result:rq2}) shows that the form is robust to small reweightings.

\subsection{TEI@$n$ Inference-Time Selection}
\label{sec:method:tei-at-n}

\begin{algorithm}[t]
\caption{\emph{TEI@$n$} Inference-Time Selection}
\label{alg:tei-at-n}
\begin{algorithmic}[1]
\STATE \textbf{Input:} Tutoring context $x$, frozen model $p_\theta$, candidate budget $N$
\STATE \textbf{Output:} Selected response $y^\star$
\FOR{$i = 1$ to $N$}
  \STATE $y_i \sim p_\theta(\cdot \mid x)$
  \STATE Extract $(V_i, \tilde M_i, Q_i, D_i)$ from $y_i$ via regex and a one-pass DTR forward hook
  \STATE $s_i \gets \alpha\, V_i + \beta\, \tilde M_i - \gamma\, Q_i + \delta\, D_i$
\ENDFOR
\STATE \textbf{return} $y^\star \gets y_{\argmax_i s_i}$
\end{algorithmic}
\end{algorithm}

Algorithm~\ref{alg:tei-at-n} generates $N$ candidates, scores each in time linear in the conversation length, and returns the best-scoring candidate. Feature extraction adds no model calls beyond the $N$ generations: $V$, $\tilde M$, and $Q$ are pure regular expressions over the trace text, and $D$ is read off the per-token JSD probe that DTR already computes during decoding.

\paragraph{Cost.} Self-consistency selection (Cons@$N$) costs $N$ generations plus an $N$-way semantic comparison; judge-based selection costs $N$ generations plus $N$ judge calls. TEI@$n$ costs $N$ generations plus $O(\text{turns})$ text operations. Empirically, TEI@8 uses $4.1\times$ the tokens of greedy decoding (16{,}334 vs.\ 3{,}984) and roughly half the cost of Cons@8 (31{,}868), while approaching the consistency-sampling improvement ceiling on the in-domain set (Section~\ref{sec:result:rq2}).

\subsection{Schoenfeld Labelling: Regex Inside TEI, GPT in Analysis}
\label{sec:method:schoenfeld}

We use two Schoenfeld~\cite{schoenfeld1985mps} labellers in this paper, separated by purpose. \emph{Inside TEI} the $V$ signal is a regex over the thinking trace with a keyword set associated with the Schoenfeld Verify episode (``let me check / verify / double-check / substitute back / \dots''), labelling a paragraph Verify when the verify-keyword hit rate exceeds a fixed threshold relative to the Explore-keyword hit rate; this preserves the no-external-model property at selection time. \emph{Outside TEI} (Section~\ref{sec:analysis:schoenfeld}) we report structural mass with a GPT-4o-mini paragraph classifier at temperature $0$, following the labelling protocol of~\citet{lee2026pedagogicalrl-thinking}. The two labellers agree on $24$ to $40\%$ of paragraphs (the regex is coarser and defaults more aggressively to General when keyword evidence is sparse); the regex is best read as a cheap proxy that empirically tracks the signal we want at deployment time. Selector-level evidence that the regex picks out higher-$V$ candidates is in Table~\ref{tab:selector-profile}.

\section{Experimental Setup and Results}
\label{sec:result}

\subsection{Setup}
\label{sec:result:setup}

\paragraph{Models.} We evaluate DeepSeek-R1-0528-Qwen3-8B~\cite{shao2024deepseekmath} (DeepSeek-R1-8B henceforth) as base and as a GRPO pedagogically-aligned variant~\cite{lee2026pedagogicalrl-thinking}. Both expose explicit thinking traces, enabling per-token DTR computation. We use temperature $0.7$, top-$p$ $0.95$, and a maximum generation length of $4{,}096$ tokens.

\paragraph{Datasets.} (i) \emph{Tutoring scenarios (in-domain)}: $1{,}000$ BigMath~\cite{albalak2025bigmath} problems of medium-to-hard difficulty per model, multi-turn dialogue against a simulated student (the aligned variant uses $500$). (ii) \emph{OpenLearnLM}~\cite{lee2026openlearnlm} (OOD): $634$ items spanning Content Knowledge ($193$), Pedagogical Knowledge ($227$), Skills ($200$), Attitude ($14$).

\paragraph{Student simulator and learning outcome.} A GPT-4o-mini student model attempts the target problem alone ($\textsc{pre\_correct}$), tutors via up to six turns with the LLM tutor, and re-attempts ($\textsc{post\_correct}$). The per-scenario learning outcome is $\Delta = \textsc{post\_correct} - \textsc{pre\_correct}$. We report $\Delta$ Solve Rate (mean over scenarios) and \emph{improvement rate} on the pre-incorrect subset (fraction with $\Delta = +1$).

\paragraph{Metrics.} Learning outcomes ($\Delta$ Solve Rate; improvement rate). Tutoring quality (LLM-judge, GPT-4o-mini): leak rate, helpful rate, hint quality (1-5), pedagogical quality (1-5). Internal: DTR, average thinking length. Selection: improvement rate of $\mathrm{Greedy}$, $\mathrm{Random@}N$, $\mathrm{DTR@}N$, $\mathrm{TEI@}N$, $\mathrm{Oracle}$. LLM-judge metrics inherit a same-model-bias risk \cite{zheng2023llmjudge}; we accordingly anchor RQ2 conclusions on $\Delta$ Solve Rate, which depends on student-model correctness rather than judge ratings. Following \cite{lee2026pedagogicalrl-thinking}, we report a composite $R_\text{total} = \tfrac{1}{3}(\Delta\mathrm{Sol} + (1 - \mathrm{Leak}) + \mathrm{Help})$.

\subsection{Main Results}
\label{sec:result:main}

Table~\ref{tab:main} reports tutoring performance for three frontier zero-shot tutors and four RL-trained variants from prior work~\cite{lee2026pedagogicalrl-thinking} alongside our three models. Frontier and RL rows are quoted verbatim from~\cite{lee2026pedagogicalrl-thinking}; our rows are measured on the conditions described in Section~\ref{sec:result:setup}.

\begin{table*}[t]
\footnotesize
\setlength{\tabcolsep}{3pt}
\centering
\caption{Main tutoring results, in-domain and out-of-distribution. GRPO alignment of the DeepSeek-R1-8B base drops $\Delta$Sol to $-.012$ and degrades every OpenLearnLM knowledge dimension. Frontier and RL rows are quoted from~\cite{lee2026pedagogicalrl-thinking}; column definitions in Section~\ref{sec:result:setup} and the table footnote. \emph{TEI@$n$} selector effects are reported separately in Tables~\ref{tab:rq2-selection} and~\ref{tab:selector-profile}.}
\label{tab:main}
\begin{tabular}{@{}llcccc|ccccc@{}}
\toprule
 & & \multicolumn{4}{c|}{In-domain} & \multicolumn{5}{c}{OpenLearnLM (OOD)} \\
Model & Method & $\Delta$Sol & Leak & Help & $R_\text{tot}$ & CK & PK & SK & Att & Avg \\
\midrule
\multicolumn{11}{l}{\emph{Frontier (zero-shot, pedagogical prompt)}} \\
\rowcolor{black!8} GPT-5.2          & Prompting & .340 & .000 & .440 & .593 & 8.08 & 8.46 & 8.63 & 8.68 & 8.46 \\
\rowcolor{black!8} Claude-4-Opus    & Prompting & .350 & .090 & .760 & .673 & 6.63 & 8.61 & 8.82 & 8.45 & 8.13 \\
\rowcolor{black!8} DeepSeek-V3.2    & Prompting & .390 & .110 & .820 & .700 & 7.46 & 7.32 & 8.63 & 8.77 & 8.05 \\
\midrule
\multicolumn{11}{l}{\emph{RL-trained baselines}} \\
\rowcolor{black!8} Qwen2.5-7B Inst.  & Unoptimized          & .120 & .300 & .180 & .333 & 7.72 & 7.14 & 7.77 & 7.57 & 7.55 \\
\rowcolor{black!8} Qwen3-8B          & PedRL Think NR       & .281 & .180 & .730 & .604 & 7.72 & 7.14 & 7.76 & 7.86 & 7.62 \\
\rowcolor{black!8} Qwen3-8B          & PedRL Think R        & .284 & .182 & .764 & .621 & 7.72 & 7.14 & 7.76 & 7.71 & 7.58 \\
\rowcolor{black!8} Qwen3-8B          & PedRL Ped Th NR      & .275 & .214 & .766 & .607 & 7.72 & 7.14 & 7.76 & 7.86 & 7.62 \\
\rowcolor{black!8} Qwen3-8B          & PedRL Ped Th R       & .294 & .172 & .776 & .633 & 7.72 & 7.14 & 7.77 & 7.86 & 7.62 \\
\midrule
\multicolumn{11}{l}{\emph{Training-Free: Comparison}} \\
DS-R1 base                & Greedy               & .180 & .583 & .220 & .272 & 2.18 & 2.20 & 7.03 & 8.07 & 4.87 \\
DS-R1 (PedRL Ped Th R)    & Greedy               & \underline{-.012} & \underline{.612} & .416 & \underline{.264} & \underline{0.62} & \underline{0.44} & 7.62 & 7.64 & \underline{4.08} \\
DS-R1 base                & DTR@4                & .180 & .581 & .224 & .274 & 2.18 & 2.20 & 7.03 & 8.07 & 4.87 \\
\midrule
\multicolumn{11}{l}{\emph{Training-Free: Ours}} \\
\rowcolor{lightblue} DS-R1 base   & \emph{TEI@4}    & \textbf{.184} & \textbf{.572} & \textbf{.228} & \textbf{.280} & 2.18 & 2.20 & 7.03 & 8.07 & 4.87 \\
\bottomrule
\end{tabular}

\vspace{2pt}
{\scriptsize Models: DS-R1 = DeepSeek-R1-0528-Qwen3-8B; DSv3.2 = DeepSeek-V3.2; the DS-R1 (PedRL Ped Th R) row is DS-R1 base aligned with the PedRL Ped Think R recipe of~\cite{lee2026pedagogicalrl-thinking}. ID: $\Delta$Sol = post-pre solve rate; Leak/Help judged by GPT-4o-mini; $R_\text{tot}=\tfrac{1}{3}(\Delta\mathrm{Sol}+(1-\mathrm{Leak})+\mathrm{Help})$. OL: CK/PK = MCQ accuracy ($0$-$10$, raw \% rescaled); SK/Att = judge rubric ($0$-$10$); Avg = mean of the four. \textbf{Bold} marks where \emph{TEI@4} is best among the four Training-Free rows; \underline{underline} marks the Greedy DS-R1 (PedRL Ped Th R) row where the alignment tax produces the worst Training-Free value.}
\end{table*}

Three observations follow from Table~\ref{tab:main}. \emph{(i) Alignment tax is sharp and negative on $\Delta$Sol.} The GRPO-aligned DeepSeek-R1-8B improves on Help ($.416$ vs.\ $.220$) and slightly on Leak relative to its base, but its $\Delta$Sol drops from $+.180$ to $-.012$. The base model is the better learning intervention despite scoring lower on Help. \emph{(ii) \emph{TEI@4} beats both DTR@4 and Greedy on the same base.} On DS-R1 base, \emph{TEI@4} raises $\Delta$Sol from $.180$ to $.184$, lowers Leak from $.583$ to $.572$, and improves $R_\text{tot}$ from $.272$ to $.280$ at no training cost; DTR@4 moves these only marginally ($+.000/-.002/+.002$). \emph{(iii) OOD pattern matches in-domain.} The aligned model loses on every OpenLearnLM knowledge dimension (CK $2.18 \to 0.62$, PK $2.20 \to 0.44$) and on the OL Avg ($4.87 \to 4.08$). Selector-level evidence beyond Table~\ref{tab:main} is reported in Tables~\ref{tab:rq2-selection} and~\ref{tab:selector-profile}.

\subsection{RQ1: DTR Does Not Predict Tutoring Quality}
\label{sec:result:rq1}

If reasoning depth transfers from mathematical accuracy to tutoring effectiveness, we should observe correlation between DTR and downstream learning outcomes. We do not. Table~\ref{tab:rq1-corr} reports the four central correlations on the DeepSeek-R1-8B base ($n = 500$ scenarios with paired DTR and judge metrics): DTR is uncorrelated with $\Delta$ Solve Rate ($r = +0.055$), with judge-rated helpfulness ($r = +0.035$), with leak rate (borderline $r = -0.076$, $p = 0.092$), and with pedagogical quality ($r = +0.051$). For comparison, the same DTR computed on the same model achieves $r = 0.683$ on math accuracy in \cite{chen2026dtr}.

\begin{table}[h]
\small
\centering
\caption{DTR is essentially uncorrelated with every tutoring outcome we measure (DeepSeek-R1-8B base, $n = 500$ scenarios with paired DTR and judge metrics; all $|r| < 0.09$).}
\label{tab:rq1-corr}
\begin{tabular}{@{}lcc@{}}
\toprule
Outcome & Pearson $r$ & $p$ \\
\midrule
$\Delta$ Solve Rate              & $+0.055$ & $0.216$ \\
helpful\_rate                    & $+0.035$ & $0.435$ \\
leak\_rate                       & $-0.076$ & $0.092$ \\
pedagogical\_quality (judge)     & $+0.051$ & $0.256$ \\
\bottomrule
\end{tabular}
\end{table}

\paragraph{Is this just a noisy measurement?} We rule out three alternative DTR formulations and demonstrate a causal intervention. \emph{(a) Layer-band DTR.} Decomposing DTR into early, middle, and late bands yields $0/24$ significant correlations after Bonferroni; a 5-fold cross-validated linear combination achieves $R^2 = -0.022$ for $\Delta$ Solve Rate. \emph{(b) Trajectory features.} Eight scenario-level DTR trajectory features produce $0/8$ significant differences between improved and non-improved groups (Mann-Whitney $p > 0.24$). \emph{(c) Difficulty estimation.} Classifying human-labelled hard versus easy Skills items yields ROC-AUC $0.513$, indistinguishable from random. \emph{(d) Causal intervention.} Consistency sampling (Cons@8) raises average DTR by $+17.3$\% ($p < 0.001$), yet tutoring quality is unchanged (Wilcoxon $p = 1.0$).

DTR is a real signal, varying systematically with task type (Skills $0.311 >$ Attitude $0.301 >$ CK $0.287 >$ PK $0.251$; ANOVA $F = 39.0$, $p < 10^{-22}$). It is simply not the signal that distinguishes a good tutor from a bad one. The TEI feature ablation in Table~\ref{tab:tei-ablation} provides a further check: dropping the DTR gate from TEI removes only $-.009$ AUC, while dropping $V$ removes $-.054$. Verification and explicit math content carry the prediction; the DTR gate adds a small bonus rather than the bulk of the signal.

\subsection{RQ2: TEI@$n$ Outperforms Alternatives}
\label{sec:result:rq2}

We compare five inference-time selection strategies on the pre-incorrect subset of the in-domain tutoring scenarios: Greedy ($N = 1$), Random@$N$, DTR@$N$, TEI@$N$, and an Oracle that selects $\Delta = +1$ when one exists. TEI@$N$ is measured via a block-grouped protocol: pre-incorrect scenarios are shuffled into blocks of size $N$; within each block we pick the scenario with the highest TEI score and check whether its $\Delta = +1$. This isolates TEI's ranking power on a fixed Greedy response distribution. We sweep $N \in \{2, 4, 8, 16\}$ with 20 random repeats per $N$.

\begin{table}[h]
\small
\centering
\caption{Improvement rate (\%) on pre-incorrect scenarios, by block-grouped selection method and $N$. 20 random repeats per $N$, DeepSeek-R1-8B base ($n = 363$ pre-incorrect).}
\label{tab:rq2-selection}
\begin{tabular}{@{}lccccc@{}}
\toprule
$N$ & Greedy & Random@$N$ & DTR@$N$ & TEI@$N$ & Oracle \\
\midrule
$2$  & 58.7 & 58.0 & 59.7 & \textbf{69.1} & \phantom{0}83.2 \\
$4$  & 58.3 & 58.6 & 61.2 & \textbf{75.7} & \phantom{0}97.5 \\
$8$  & 59.0 & 58.1 & 66.8 & \textbf{81.9} & \phantom{0}99.8 \\
$16$ & 58.6 & 59.1 & 72.5 & \textbf{84.8} & 100.0 \\
\bottomrule
\end{tabular}
\end{table}

At $N = 8$, TEI@$N$ reaches $81.9\%$ versus Greedy $59.0\%$ ($+22.9$pp). The gap is stable across $N$; DTR@$N$ trails TEI@$N$ by $13$ to $15$pp at every $N$, and Random@$N$ never exceeds Greedy.

\paragraph{What does TEI actually pick?} Table~\ref{tab:selector-profile} reports the mean surface features of TEI@4-selected candidates versus the Greedy ($N{=}1$) candidate on the same $500$ in-domain scenarios. The selector raises Verify-keyword ratio by $+.243$ and the (capped, raw) math-step count by $+1.91$ per turn, while reducing the ends-question rate by $-.026$: the exact directions the three signed weights reward. Across all $500$ scenarios, TEI picks a non-Greedy candidate $70.0\%$ of the time (Greedy is selected $30.0\%$), with each of the three non-Greedy candidates chosen in $22{-}24\%$ of cases; the mean candidate-pool TEI gap (max $-$ min) is $0.70$, showing the four candidates differ enough for selection to be non-trivial. The selector therefore is not just re-ranking near-duplicates: it pulls the response toward more verification-rich, more math-explicit, and less terminal-question content, consistent with the signed weights in Eq.~\eqref{eq:tei}.

\begin{table}[h]
\small
\centering
\caption{Selector surface profile ($n{=}500$ in-domain scenarios, DeepSeek-R1-8B base). \emph{TEI@4} shifts the three signed signals in the rewarded directions; the gate $D$ is flat in this pool (no candidate exceeds $\mathrm{DTR}\!\geq\!0.4$).}
\label{tab:selector-profile}
\begin{tabular}{@{}lccccc@{}}
\toprule
Cell & $V$ & $\tilde M$ & $M_\text{raw}$ & $Q$ & $D{=}\mathbf{1}[\mathrm{DTR}\!\geq\!.4]$ \\
\midrule
Greedy ($N{=}1$)     & .497 & .574 & 11.96 & .031 & .000 \\
Random@4 (pool mean) & .488 & .575 & 11.99 & .033 & .000 \\
\emph{TEI@4} (selected) & \textbf{.740} & \textbf{.657} & \textbf{13.87} & \textbf{.005} & .000 \\
Worst-TEI            & .240 & .478 & \phantom{0}9.85 & .073 & .000 \\
\midrule
$\Delta$ TEI@4 $-$ Greedy & $+.243$ & $+.084$ & $+1.91$ & $-.026$ & $.000$ \\
\bottomrule
\end{tabular}
\end{table}

\paragraph{Feature contribution.} We evaluate TEI's pool-ranking by AUC against the binary $\Delta = +1$ outcome (Table~\ref{tab:tei-ablation}). Full TEI beats every single-feature baseline ($.698$ vs.\ best single $.637$). Leave-one-out drops are largest for $V$ ($-.054$) and $\tilde M$ ($-.036$); $Q$ and the DTR gate $D$ each contribute under $.01$ AUC, consistent with RQ1. An Explore-keyword term, if added, also moves AUC by under $.005$: regex Verify and Explore partition each thinking trace into anti-correlated blocks ($\rho(V, E) = -0.996$ on $n = 363$), so an Explore term acts as a proxy for the absence of Verify rather than an independent signal.

\begin{table}[h]
\small
\centering
\caption{\emph{TEI} feature ablation on DeepSeek-R1-8B base. \emph{TEI} beats every single-feature baseline; $V$ is the most important component; the DTR gate $D$ contributes the smallest amount. Pre-incorrect subset ($n = 363$).}
\label{tab:tei-ablation}
\begin{tabular}{@{}lcc@{}}
\toprule
Scorer & AUC & $\rho$ \\
\midrule
Random                     & .442 & n/a     \\
$V$ alone                  & .636 & $\phantom{-}.242$ \\
$\tilde M$ alone           & .637 & $\phantom{-}.233$ \\
$Q$ alone                  & .476 & $-.104$ \\
DTR (continuous)           & .525 & $\phantom{-}.046$ \\
$D{=}\mathbf{1}[\mathrm{DTR}\!\geq\!0.4]$ & .530 & $\phantom{-}.118$ \\
\midrule
\textbf{TEI (full)}        & \textbf{.698} & $\phantom{-}.338$ \\
TEI $+$ Explore (no $D$)   & .682 & $\phantom{-}.310$ \\
w/o $V$                    & .644 & $\phantom{-}.247$ \\
w/o $\tilde M$             & .663 & $\phantom{-}.288$ \\
w/o $Q$                    & .690 & $\phantom{-}.324$ \\
w/o $D$                    & .689 & $\phantom{-}.322$ \\
\bottomrule
\end{tabular}
\end{table}

\subsection{RQ3: Alignment Tax}
\label{sec:result:rq3}

Table~\ref{tab:main} already documents the $\Delta$Sol collapse from $+.180$ to $-.012$ under GRPO alignment. We now decompose the tax across response geometry, OpenLearnLM accuracy, and tutoring-quality components.

\begin{table}[h]
\small
\centering
\caption{Alignment tax decomposition (DeepSeek-R1-8B base vs.\ GRPO-aligned). Gains on surface helpfulness; losses on thinking length, $\Delta$Sol, and OL CK/PK.}
\label{tab:rq3-alignment-tax}
\begin{tabular}{@{}lccc@{}}
\toprule
Metric & Base & GRPO-aligned & $\Delta$ \\
\midrule
Avg thinking length (words/turn)   & 1{,}764 & 119   & $-93.3\%$ \\
Avg visible length (words/turn)    & 194     & 21    & $-89.3\%$ \\
$\Delta$ Solve Rate (all scenarios)& $+.180$ & $-.012$ & $-.192$ \\
Tutoring helpful\_rate             & .220    & .416  & $+.196$ \\
Tutoring leak\_rate                & .583    & .612  & $+.029$ \\
Tutoring hint\_quality (1-5)       & 4.73    & 3.53  & $-1.20$ \\
Tutoring pedagogical\_quality (1-5)& 4.11    & 3.66  & $-0.45$ \\
OpenLearnLM CK (0-10)              & 2.18    & 0.62  & $-71\%$ rel.\ \\
OpenLearnLM PK (0-10)              & 2.20    & 0.44  & $-80\%$ rel.\ \\
OpenLearnLM Skills (0-10)          & 7.03    & 7.62  & $+0.59$ \\
OpenLearnLM Avg (0-10)             & 4.87    & 4.08  & $-0.79$ \\
\bottomrule
\end{tabular}
\end{table}

The pattern is unambiguous. Pedagogical alignment cuts thinking length by an order of magnitude and visible length by nearly the same amount; the visible mathematical content density drops from $25.5\%$ to $9.5\%$; the student's post-tutoring solve rate actually falls below the pre-tutoring rate ($\Delta = -.012$). The only metric that improves is judge-rated helpfulness, and judge-rated hint and pedagogical quality also fall. The aligned model is rated as more helpful by the LLM judge but is, by the learning-outcome measure, actively harmful on average. We interpret this as evidence that pedagogical RL with helpfulness-style rewards can optimize a proxy that is decoupled from learning.

\subsection{RQ4: Out-of-Distribution Validation}
\label{sec:result:rq4}

The OpenLearnLM columns of Table~\ref{tab:main} confirm the in-domain alignment-tax pattern on an independent benchmark. The aligned model loses on Content Knowledge ($2.18 \to 0.62$) and Pedagogical Knowledge ($2.20 \to 0.44$) and on the four-dimension Avg ($4.87 \to 4.08$). The one OL dimension that rewards the aligned model, Skills, is also the dimension whose judge rubric agrees most closely with the helpfulness reward GRPO optimised for, a consistency-with-its-own-reward result rather than a transfer one.

\section{Analysis}
\label{sec:analysis}

We turn from selection outcomes (Section~\ref{sec:result}) to \emph{what the tutor is actually saying} during a dialog, using a behavioural codebook approach. We label every tutor sentence (after stripping the thinking block) with one or more of an 82-code educational codebook organised into 7 major categories: Mathematical Problem Solving (MPS), Mathematical Knowledge for Teaching (MKT), Cognition, Metacognition, Pedagogical Intent Utterance (PIU), Student Intent Utterance, and Interaction/Discourse. Labels are produced by GPT-4o-mini at temperature $0$ with the codebook prompt inlined; no human coding. The codebook run covers $119{,}009$ tutor sentences across $3{,}500$ scenarios. The tables and figures below report three cells: Base Greedy, Aligned GRPO, and Base \emph{TEI@4}-selected candidates. A parallel Schoenfeld trace analysis (Section~\ref{sec:analysis:schoenfeld}) covers the same cells.

\subsection{Major Category Distribution}
\label{sec:analysis:cat}

Table~\ref{tab:cat-4} reports sentence-multilabel percentages over a four-category reduction.

\begin{table}[h]
\small
\centering
\caption{Codebook category distribution (sentence-multilabel \%, four-category view). The GRPO-aligned cell drops MKT by nearly two thirds, nearly triples Interaction, and ramps PIU; selectors on the base shift category mass only marginally.\protect\footnotemark[1]}
\label{tab:cat-4}
\begin{tabular}{@{}lcccc@{}}
\toprule
Cell & MPS & MKT & PIU & Interaction \\
\midrule
Base Greedy        & 43.1 & 18.1 & 20.0 & \phantom{0}4.4 \\
Base \emph{TEI@4}  & 41.4 & 20.9 & 19.6 & \phantom{0}3.8 \\
Aligned (GRPO)     & 24.7 & \textbf{6.5} & \textbf{28.7} & \textbf{12.3} \\
\bottomrule
\end{tabular}
\footnotetext[1]{MPS = Mathematical Problem Solving; MKT = Mathematical Knowledge for Teaching; PIU = Pedagogical Intent Utterance; Interaction = Cognition + Metacognition + Student-Intent + Interaction/Discourse.}
\end{table}

Two patterns are striking. First, the aligned model's MKT rate collapses from $18.1\%$ to $6.5\%$ (a $64\%$ relative drop) while its Interaction rate nearly triples ($4.4\%$ to $12.3\%$) and its PIU rate climbs from $20.0\%$ to $28.7\%$. The same direction is reported by \citet{lee2026pedagogicalrl-thinking} for GRPO-aligned tutors: the policy shifts away from declarative mathematical content toward interactive, pedagogical-intent utterances. Second, \emph{TEI@4} selection on the base model leaves all four category rates within $1.7$pp of the base Greedy values, while pulling MKT slightly higher ($18.1\% \to 20.9\%$); the selector shifts the candidate pool toward more declarative mathematical content without redistributing category mass at the macroscopic level.

\subsection{Core Code Frequency}
\label{sec:analysis:core6}

Figure~\ref{fig:behaviour} (\emph{left}, panel b) plots the rate of six pedagogically-relevant codes per cell. The base model's most frequent codes are content-delivering: \emph{Concept explanation}, \emph{Information provision}, \emph{Step-by-step instruction}, and \emph{Praise}. The aligned model shifts the head of the distribution toward interactional codes: \emph{Step-by-step instruction} survives but is joined by \emph{Exploratory question} and \emph{Open-ended question} at substantially higher rates. The Exploratory-question rate alone increases roughly fourfold under alignment, mirroring the per-turn Schoenfeld Explore shift in Section~\ref{sec:analysis:schoenfeld}.

\subsection{Polya Phase Coverage}
\label{sec:analysis:polya}

\begin{figure*}[t]
\centering
\begin{minipage}[b]{0.60\textwidth}
  \centering
  \includegraphics[width=\linewidth]{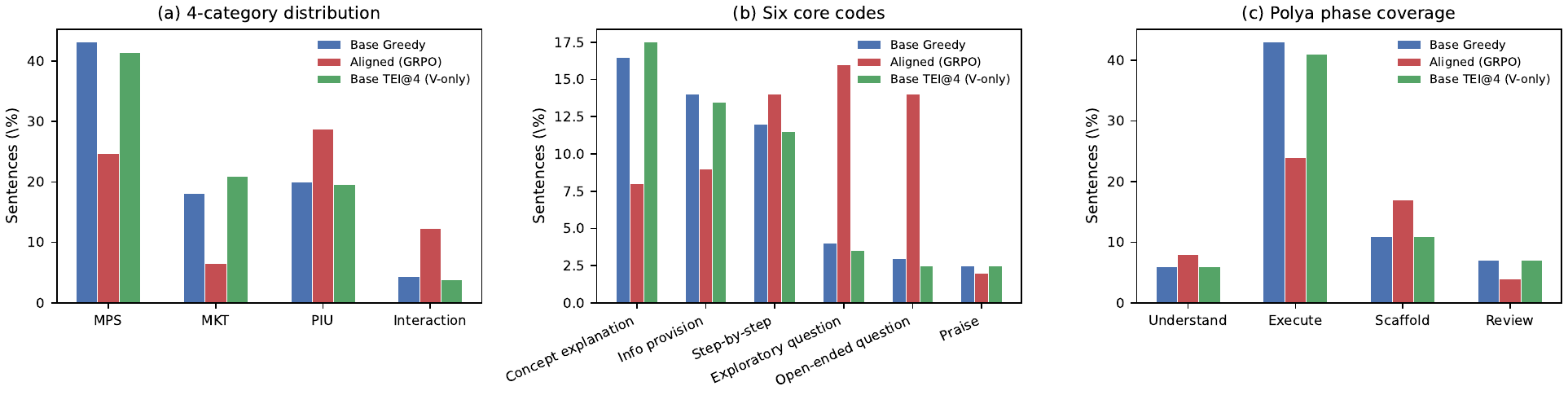}
\end{minipage}\hfill
\begin{minipage}[b]{0.38\textwidth}
  \centering
  \includegraphics[width=\linewidth]{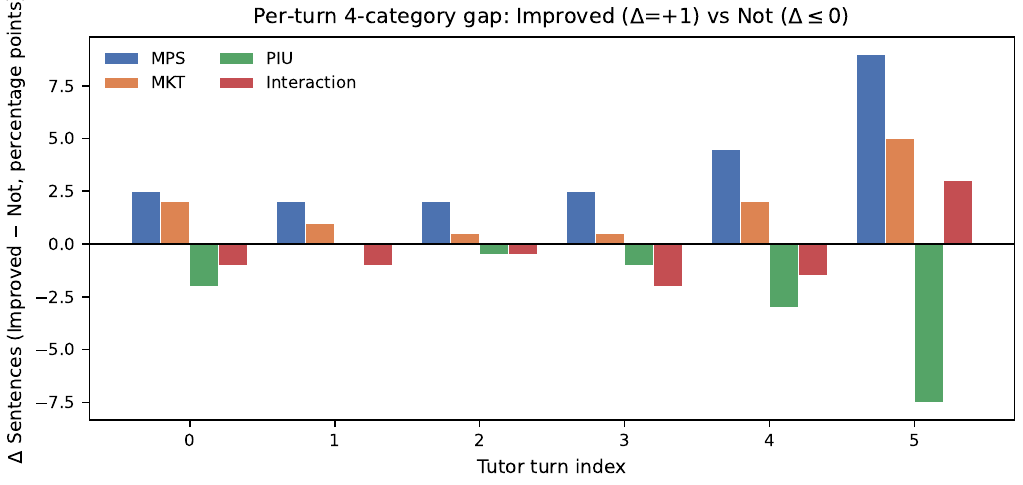}
\end{minipage}
\caption{Codebook behavioural analysis. (\emph{left}) On $119{,}009$ tutor sentences: (a) four-category distribution per cell, (b) six core codes per cell, (c) Polya phase coverage per cell. Labels by GPT-4o-mini at temperature $0$. (\emph{right}) Per-turn four-category gap (pp) between improved ($\Delta = +1$) and not-improved ($\Delta \leq 0$) dialogs on DeepSeek-R1-8B base.}
\label{fig:behaviour}
\end{figure*}

We map each fine code to one of Polya's phases~\cite{polya1945howto} and report phase coverage in Figure~\ref{fig:behaviour} (\emph{left}, panel c). Execute dominates every cell; the aligned cell loses Execute (the MKT mass that drops) and gains Scaffold (the question-heavy PIU shift), while \emph{TEI@4}-selected candidates retain the base distribution.

\subsection{Success vs.\ Failure Dialogs}
\label{sec:analysis:success}

Joining the codebook labels with the per-scenario $\Delta$ Solve outcome (Section~\ref{sec:result:setup}), Figure~\ref{fig:behaviour} (\emph{right}) reports the per-turn percentage-point gap (improved $-$ not improved) for the four major categories on the base model. MKT and MPS are more frequent at later turn positions in successful dialogs, while PIU and Interaction flip sign in some turn positions, consistent with the aligned-model signature being over-represented among the failures rather than among the successes.

\subsection{Alignment Tax in the Raw Stream}
\label{sec:analysis:alignment-tax}

\begin{figure*}[t]
\centering
\begin{minipage}[b]{0.58\textwidth}
  \centering
  \includegraphics[width=\linewidth]{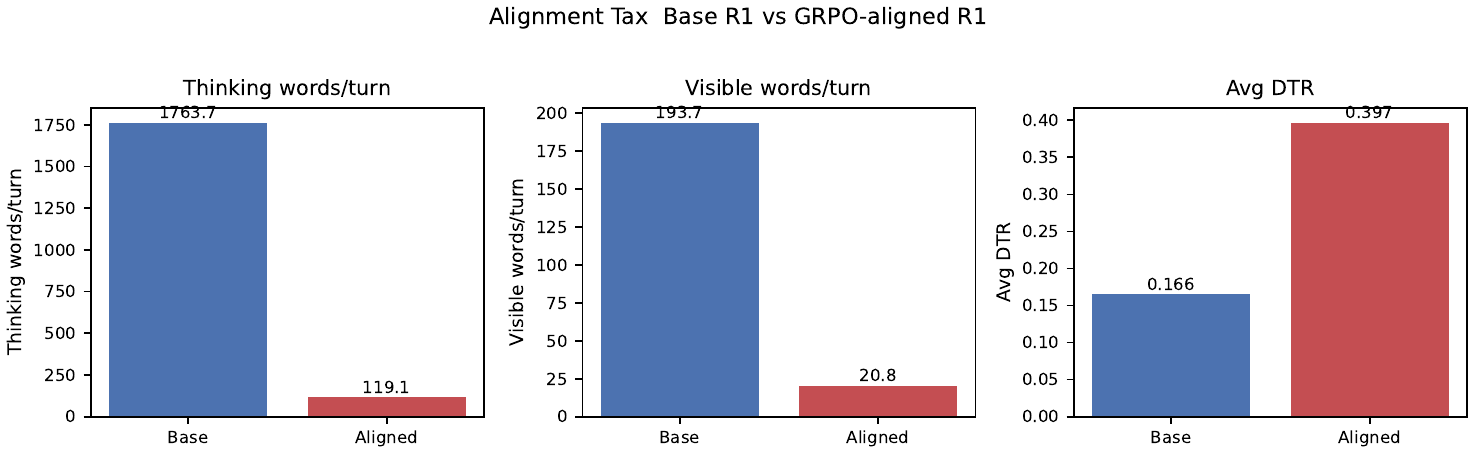}
\end{minipage}\hfill
\begin{minipage}[b]{0.40\textwidth}
  \centering
  \includegraphics[width=\linewidth]{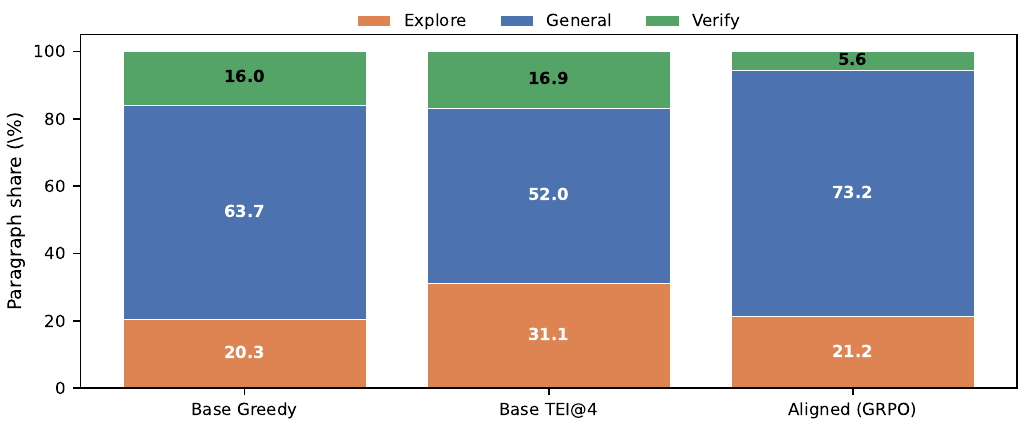}
\end{minipage}
\caption{Thinking-trace analysis. (\emph{left}) Alignment tax in the raw stream: thinking words/turn, visible words/turn, and average DTR, on Base R1 vs.\ GRPO-aligned R1. Aligned (GRPO) collapses thinking by an order of magnitude; \emph{TEI@4} on the base shifts these only marginally and is not shown to keep the panel readable. (\emph{right}) Schoenfeld phase distribution (GPT-4o-mini classifier) in tutor thinking-trace paragraphs, three cells. GRPO halves Verify and inflates General; \emph{TEI@4} on the base lifts Explore by $+10.8$pp into a deeper deliberation density.}
\label{fig:traces}
\end{figure*}

The codebook-level signature has a direct counterpart in the raw response statistics (Figure~\ref{fig:traces}, \emph{left}; numbers also in Table~\ref{tab:rq3-alignment-tax}). The aligned model's thinking length collapses from $1{,}764$ to $119$ words per turn ($-93.3\%$), and its visible response from $194$ to $21$ words ($-89.3\%$). This is the empirical face of the alignment tax: GRPO pedagogical alignment does not just rephrase the tutor's surface, it strips the visible mathematics and the internal deliberation that produced it. \emph{TEI@4} selection on the base shifts these raw-length averages only modestly (thinking $-3\%$, visible $+12\%$) while delivering the improvement-rate gain reported in Table~\ref{tab:rq2-selection}; the structural shifts \emph{within} those traces are documented in Section~\ref{sec:analysis:schoenfeld}.

\subsection{Schoenfeld Phase Distribution in Thinking Traces}
\label{sec:analysis:schoenfeld}

The codebook analysis above operates on the tutor's \emph{visible} output. We complement it with a Schoenfeld~\cite{schoenfeld1985mps} phase analysis of the tutor's \emph{thinking trace}, labelling every paragraph inside the thinking block with the same GPT-4o-mini structural classifier used for the codebook (temperature $0$, one shot per paragraph). The regex variant used inside the $V$ signal of \emph{TEI} (Section~\ref{sec:method:schoenfeld}) is faster but coarser; we use the GPT classifier here so that the analysis section reports cross-cell structural mass that the regex would underdistinguish at the General default.

\begin{table}[h]
\small
\centering
\caption{Schoenfeld phase distribution in thinking-trace paragraphs (\%, GPT-4o-mini paragraph classifier). GRPO compresses the trace and halves Verify mass; \emph{TEI@4} selection shifts mass out of General into Explore, reaching higher deliberation density on the same base.}
\label{tab:schoenfeld}
\begin{tabular}{@{}lrrrr@{}}
\toprule
Cell & $n$ paragraphs & Explore & General & Verify \\
\midrule
Base Greedy        & 2{,}335 & 20.3 & 63.7 & 16.0 \\
Base \emph{TEI@4}  & 1{,}175 & \textbf{31.1} & 52.0 & 16.9 \\
Aligned (GRPO)     & 4{,}235 & 21.2 & \textbf{73.2} & \underline{\phantom{0}5.6} \\
\bottomrule
\end{tabular}
\end{table}

Two observations follow. \emph{(i) Alignment compresses the trace and shifts Verify mass into the General default.} The aligned cell's Verify rate drops to $5.6\%$ (from $16.0\%$ Base R1), while General rises to $73.2\%$. GRPO compresses the trace to ${\sim}119$ words per turn (Table~\ref{tab:rq3-alignment-tax}); a shorter trace contains less recognisable verification scaffolding and more procedural sentences that the classifier categorises as General. \emph{(ii) \emph{TEI@4} selects deliberation-rich candidates.} Starting from the Base R1 candidate pool, \emph{TEI@4} raises Explore by $+10.8$pp ($20.3\% \to 31.1\%$) and lowers General by $-11.7$pp, reaching a deeper deliberation density without changing the model. Verify itself is roughly preserved ($16.0 \to 16.9\%$): the regex-$V$ signal that \emph{TEI} rewards (Table~\ref{tab:selector-profile}, $V$ ratio $.497 \to .740$) correlates broadly with deliberation-heavy thinking rather than narrowly with the Verify episode the GPT classifier identifies.

\section{Discussion \& Conclusion}
\label{sec:conclusion}

Mathematical accuracy and tutoring effectiveness inhabit different axes~\cite{chi2014icap}: the former is token-level correctness, which DTR measures well, the latter is turn-level interaction structure, which DTR is uninformative about. Our four-angle null result on DTR (correlation, causal steering, layer-band, trajectory) is therefore not a measurement failure but a category-level mismatch, and \emph{TEI} is designed to target the right axis: a Schoenfeld-Verify keyword regex captures how the tutor's hidden reasoning is organised, the surface math-step density captures how much mathematical content the tutor actually delivers, the ends-question rate penalises Socratic over-questioning, and a deep-reasoning gate adds a small bonus when the DTR probe indicates sustained layer-wise deliberation. The \emph{TEI@$n$} selector helps most when the candidate pool is diverse and the base distribution is well-balanced; the gain is consistent across $N$ (Table~\ref{tab:rq2-selection}), and the selector surface profile in Table~\ref{tab:selector-profile} confirms that the $V$ signal really does pick out candidates with more verification-keyword paragraphs (Verify-keyword ratio $.497 \to .740$).

The most consequential finding is the alignment tax. GRPO pedagogical tuning improves what an LLM judge rewards (Help $+.196$, Table~\ref{tab:rq3-alignment-tax}) and what a Socratic-tutoring rubric would credit (Exploratory-question rate rises sharply, Table~\ref{tab:cat-4}). It simultaneously degrades what the student's post-tutoring solve rate measures ($-.192$ in $\Delta$Sol), the thinking trace that drives mathematical correctness ($-93.3\%$ in length), the declarative mathematical content delivered to the student (MKT $-11.6$pp), and OpenLearnLM CK and PK accuracy ($-71\%$ and $-80\%$ relative). The aligned model has been optimised toward a proxy that is decoupled from learning. For practitioners, the takeaway is that a useful pedagogical alignment lever sits between full RL retraining and unprincipled greedy decoding: inference-time selection with a tiny interpretable index retains the base model's mathematical depth and exposes four interpretable weights for behavioural tuning, where full RL alignment is currently a sharper but more dangerous instrument.

In summary, we (C1) propose \emph{TEI}, a training-free judge-free four-signal index that combines a Schoenfeld-Verify regex, a math-step density, an ends-question rate, and a DTR gate; (C2) report \emph{TEI@$n$} as an inference-time selector that lifts the improvement rate on pre-incorrect scenarios by $+22.9$pp at $N = 8$; (C3) reproduce an 82-code educational codebook with a one-shot GPT-4o-mini classifier on $119{,}009$ tutor sentences, supplying the behavioural grounding for the analysis; and (C4) measure a multi-dimensional alignment tax in pedagogical RL.

\paragraph{Limitations.} Five caveats qualify these findings. (i) $\Delta$ Solve Rate uses GPT-4o-mini as a synthetic learner rather than real students. (ii) All experiments are in mathematics, so \emph{TEI} weights and signals may need re-derivation for other subjects. (iii) Helpful and leak rates are LLM-judge metrics and inherit same-model-bias risk; the primary outcome ($\Delta$ Solve Rate) does not. (iv) The Schoenfeld $V$ signal is regex-derived and surface-level: a deeper GPT-4o-mini paragraph classifier disagrees with it on more than half of paragraphs (Section~\ref{sec:method:schoenfeld}), so the regex is best read as a cheap proxy that empirically tracks the signal we want at deployment time. (v) The four \emph{TEI} weights are fixed \emph{a priori} from theory; domain calibration could improve performance at the cost of re-introducing a labelled-outcome requirement.

\paragraph{Future work.} Validation with human learners; extension to non-math domains; constrained calibration of the four \emph{TEI} weights against a small held-out outcome while preserving the judge-free property at deployment time.


\section*{Use of Generative AI}
\label{sec:genai}

Generative AI tools were used in this work as follows. \emph{Writing.} Claude (Anthropic) and ChatGPT (OpenAI) assisted with proofreading and copy-editing of the manuscript; all experimental design, analysis decisions, and final wording were made by the authors. \emph{Code.} The same tools provided code scaffolds for data extraction, labelling, and figure generation; every script was reviewed and modified by the authors before use. \emph{Pipeline.} GPT-4o-mini served as the simulated student in the $\Delta$ Solve Rate evaluation, as the LLM judge for leak / helpful / hint / pedagogical-quality metrics, and as a structure-only classifier for Schoenfeld episode labelling and 82-code codebook labelling of tutor sentences. No AI was used to produce final claims or numbers without author verification.




\end{document}